\documentclass[twocolumn,prl,preprintnumbers,amsmath,amssymb,superscriptaddress]{revtex4-1}

\usepackage[draft]{hyperref}
\usepackage{amsmath}
\usepackage{amssymb}
\usepackage{bm}
\usepackage{graphicx}          
\usepackage{dcolumn}
\usepackage{color}
\usepackage{epstopdf}

\usepackage[normalem]{ulem}

\newcommand{\red}{\textcolor{black}}
\newcommand{\blue}{\textcolor{blue}}
\newcommand{\ba}{\begin{array}}
\newcommand{\ea}{\end{array}}
\newcommand{\be}{\begin{equation}}
\newcommand{\ee}{\end{equation}}
\newcommand{\bea}{\begin{eqnarray}}
\newcommand{\eea}{\end{eqnarray}}
\newcommand{\bfv}{{\bf v}}

\newcommand{\bfk}{{\bf k}}
\newcommand{\bfr}{{\bf r}}
\newcommand{\zobs}{{z_{\rm obs}}}
\newcommand{\mobius}{{{M}{\"o}bius}}

\begin{document}

\title{Observation of polarization singularities and topological textures in sound waves}

\author{Ruben D. Muelas-Hurtado}
\address{School of Civil and Geomatic Engineering, Universidad del Valle, 760032 Cali, Colombia}
\address{School of Mechanical Engineering, Universidad del Valle, 760032 Cali, Colombia}

\author{Karen Volke-Sep{\'u}lveda}
\address{Instituto de F\'isica, Universidad Nacional Aut\'onoma de M\'exico, Cd. de M\'exico, C.P. 04510, M\'exico}

\author{Joao L. Ealo}
\address{School of Mechanical Engineering, Universidad del Valle, 760032 Cali, Colombia}
\address{Centro de Investigaci\'on e Innovaci\'on en Bioinform\'atica y Fot\'onica, Universidad del Valle, 760032 Cali, Colombia}

\author{Franco~Nori}
\address{Theoretical Quantum Physics Laboratory, RIKEN Cluster for Pioneering Research, Wako-shi, Saitama 351-0198, Japan}
\address{Physics Department, University of Michigan, Ann Arbor, Michigan 48109-1040, USA}

\author{Miguel A. Alonso}
\address{Aix Marseille Univ, CNRS, Centrale Marseille, Institut Fresnel, Marseille, France}
\address{The Institute of Optics, University of Rochester, Rochester, New York 14627, USA}

\author{Konstantin Y. Bliokh}
\address{Theoretical Quantum Physics Laboratory, RIKEN Cluster for Pioneering Research, Wako-shi, Saitama 351-0198, Japan}

\author{Etienne Brasselet}
\address{Univ. Bordeaux, CNRS, LOMA, UMR 5798, F-33400 Talence, France}

\begin{abstract}
 Polarization singularities and topological polarization structures are generic features of inhomogeneous vector wave fields of any nature. However, their experimental studies mostly remain restricted to optical waves. Here we report observation of polarization singularities,  topological {\mobius}-strip structures and skyrmionic textures in 3D polarization fields of inhomogeneous sound waves. Our experiments are made in the ultrasonic domain using nonparaxial propagating fields generated by space-variant 2D acoustic sources. We also retrieve distributions of the 3D spin density in these fields. Our results open the avenue to investigations and applications of topological features and nontrivial 3D vector properties of structured sound waves. 
\end{abstract}

\maketitle

{\it Introduction.---}
Polarization is an inherent property of monochromatic vector waves, which describes the trajectory of the wave field over its oscillation period. Polarization is routinely used for transverse (e.g., electromagnetic or elastic shear) waves, where it lies in the plane orthogonal to the wavevector {\bf k} for a single plane wave and is also responsible for the spin of the wave \cite{Azzam_book,Andrews_book}. For longitudinal (e.g., sound) waves, the vector field is collinear with ${\bf k}$ for a single plane wave, and it might seem that the polarization properties of such waves are trivial. However, the interference of multiple longitudinal plane waves with different wavevectors ${\bf k}_j$, $j=1,...,N$, causes the polarization at a given point ${\bf r}$ to become an ellipse with arbitrary 3D orientation and ellipticity, and hence to have spatially-varying spin density \cite{Jones1973,Shi2019,Bliokh2019_II,Bliokh2022}. Thus, polarization properties of {\it structured} vector waves share similar generic features independently of the transverse (divergence-free) or longitudinal (curl-free) character of the wave field.

Complex inhomogeneous fields can be characterized via their {\it singularities} and {\it topological structures}. Polarization singularities (e.g., the C-points or C-lines where polarization is circular) have been analyzed in detail for optical fields \cite{Nye1987,BerryDennis2001,Dennis2002,Flossmann2005,Angelis2018,Dennis2009,BAD2019,Wang2021APL}, and it was found that the orientation of the 3D polarization ellipse in a vicinity of a generic (non-degenerate) C-point has a nontrivial {\it M\"{o}bius-strip} topology \cite{Freund2010,Dennis2010,Bauer2015,Galvez2017,Bauer2019,Tekce2019,BAD2019, Wang2021APL}. Recently, some of us argued that the same topological structures can appear naturally in inhomogeneous sound waves \cite{Bliokh2021} and the present work is in a line of emerging studies of the vector properties of sound waves. First, nonzero spin densities in structured sound waves have recently attracted great attention and offered novel methods of contactless manipulation of objects with sound waves \cite{Shi2019,Bliokh2019_II,Toftul2019,Long2020,Long2020_II,Wei2020,Wang2021}. Second, skyrmionic polarization textures have been recently described and observed in both optical \cite{Beckley2010,Donati2016,Tsesses2018,Du2019,Gao2020,Sugic2021} and sound \cite{Ge2021} waves. Finally, polarization knots were recently analyzed for both optical and sound waves \cite{Laroque2018,Pisanty2019,Sugic2020,Ferrer-Garcia2021}. It is worth mentioning that other types of classical waves (e.g., water waves) can also exhibit similar polarization-related features \cite{Bliokh2022,Sugic2020,Bliokh2021}.

Here we report on the first observation of polarization singularities, polarization M\"{o}bius strips and polarization skyrmionic textures in structured propagating sound waves. We generate nonparaxial Bessel-like sound beams with a pair of C-points near the beam center, and also investigate three-wave interference with a periodic lattice of C-points. We reconstruct 3D polarization-ellipse distributions of the acoustic velocity field, observe the M\"{o}bius-strip topology around the C-points, and also retrieve the 3D spin density and identify skyrmionic texture in the propagating field. Our results demonstrate that sound waves exhibit the same rich variety of topological polarization features as optical waves. 

\begin{figure}[b!]
\centering\includegraphics[width = 0.95\columnwidth]{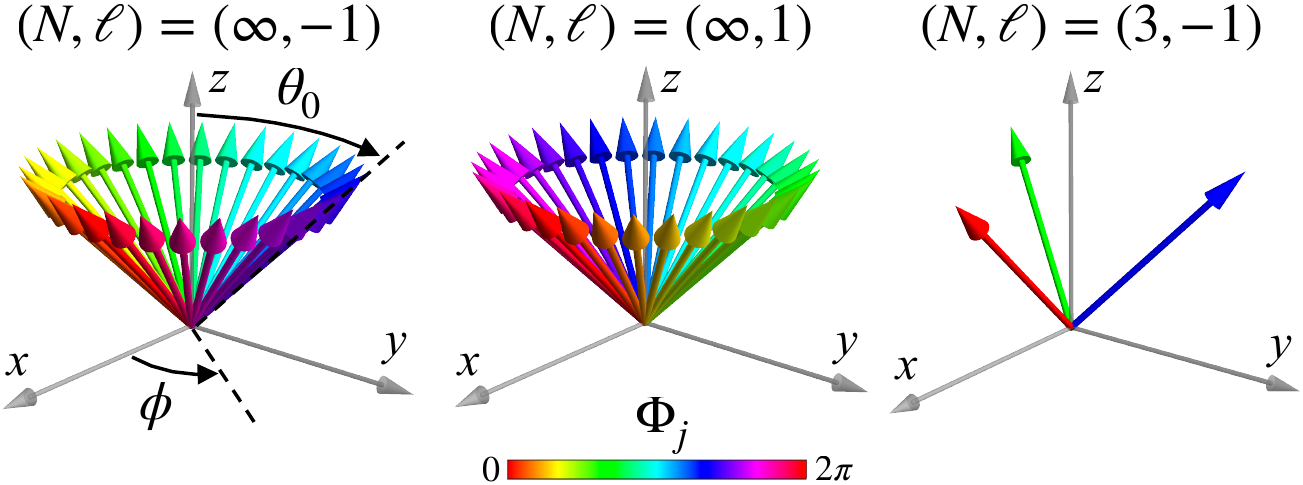}
\caption{Wavevectors of $N$ interfering planes waves, Eqs.~(\ref{eq1}) and (\ref{eq2}), evenly distributed over the cone with the polar angle $\theta_0 = \pi/4$ and having relative vortex phases $\Phi_j = \ell \phi$ (color-coded). The three cases shown here correspond to the Bessel beams with $\ell=\pm 1$ and a three-wave interference with $\ell = -1$.}
\label{Fig1}
\end{figure}

\begin{figure}[t!]
\centering\includegraphics[width = 0.8\columnwidth]{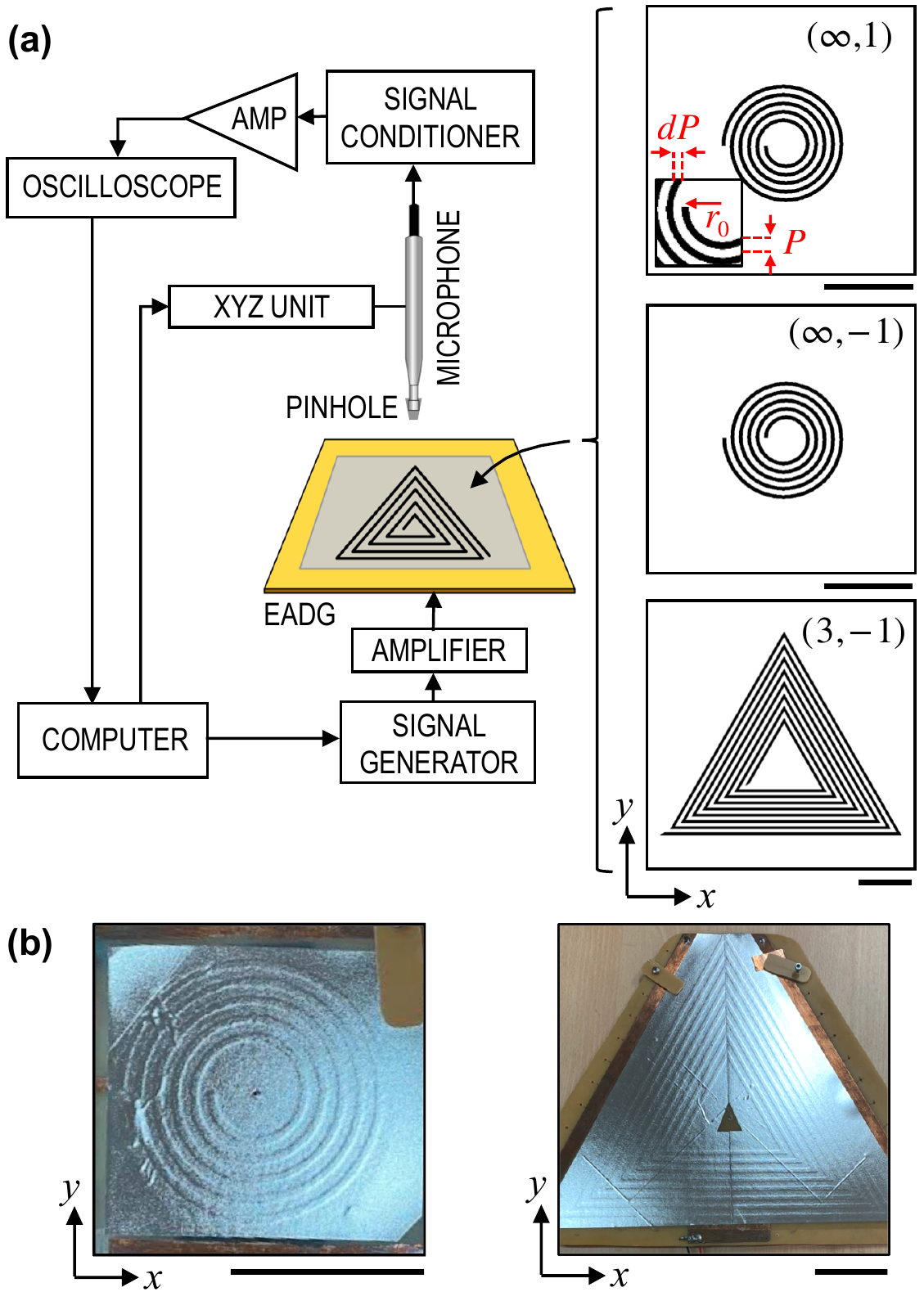}
\caption{(a) Sketch of the experimental setup involving 2D structured electro-active diffraction gratings (EADG). (b) Top view of the fabricated acoustic sources corresponding to the Bessel beam with $(N,\ell) = (\infty,1)$ and three-wave interference $(N,\ell) = (3,-1)$. Scale bars: 5~cm.}
\label{Fig2}
\end{figure}

{\it Sound waves under consideration.---}
We consider monochromatic propagating sound waves that can be described by two distinct complex space-variant fields: the scalar pressure field $p ({\bf r})$ and the vector velocity field ${\bf v} ({\bf r}) = - (i/\rho\omega) {\bm \nabla} p ({\bf r})$, where $\rho$ is the mass density of the medium (air, in our case) and $\omega$ is the angular frequency. The real time-dependent fields are obtained by applying ${\rm Re}\! \left[...\, e^{-i\omega t}\right]$ to the complex fields $p$ and ${\bf v}$ and polarization is described by the velocity field, which can thus be deduced from the gradient of the pressure field.

We consider inhomogeneous sound waves resulting from the superposition of $N$ plane waves with equal amplitudes and wavevectors evenly distributed over the circle $k_z = k \cos \theta_0$, $k = \omega / c$, Fig.~\ref{Fig1}, where $c$ is the speed of sound. The pressure and velocity fields of this superposition can be written as \cite{Bliokh2021}
\bea
\label{eq1}
p = v_0 \rho c \sum_{j=1}^{N} \exp\!\left( i\, {\bfk}_j\cdot \bfr + i \, \Phi_j \right)\,,\\
\label{eq2}
\bfv = v_0 \sum_{j=1}^{N} \bar{\bfk}_j \exp\!\left( i\, {\bfk}_j\cdot \bfr + i \, \Phi_j \right)\,,
\eea
where $v_0$ is the common real-valued velocity amplitude of every constituting plane wave, and $ \bar{\bfk}_j \equiv {\bf k}_j/k = (\sin\theta_0 \cos\phi_j , \sin\theta_0 \sin\phi_j , \cos\theta_0)$ with $\phi_j = 2\pi(j-1)/N$ are the directions of the wavevectors, using spherical angles $(\theta,\phi)$. In addition, we choose the relative phases of the interfering waves to correspond to the vortex with an integer topological charge $\ell$, i.e., $\Phi_j = \ell \phi_j$. Specifically, in our experiments we generate fields [Eqs.~(\ref{eq1}) and (\ref{eq2})] with $N=\infty$ (Bessel beams) and $N=3$ (three-wave interference), topological charges $\ell = \pm 1$, and $\theta_0 = \pi /4$.

The polarization singularities of the vector field ${\bf v}$ are the C-points (in the 2D $(x,y)$-plane) or C-lines (in 3D space) \cite{Nye1987,BerryDennis2001,Dennis2002,Flossmann2005,Angelis2018,Dennis2009,BAD2019,Wang2021APL}, which correspond to phase singularities (vortices) of the quadratic scalar field ${\bf v}\cdot {\bf v}$. \red{We note that, generally, polarization singularities do not coincide with phase singularities of the pressure field.} As predicted in Ref.~\citenum{Bliokh2021}, a Bessel beam with $\ell = \pm 1$ has a double-degenerate C-point at its centre, which can be split into a pair of non-degenerate C-points by a perturbation breaking the cylindrical symmetry of the field \cite{Bliokh2008,Vyas2013,Otte2018}, whereas the three-wave interference generates a periodic lattice of non-degenerate C-points. The orientation of the major axes of the 3D polarization ellipses has a M\"{o}bius-strip topology along a closed contour embracing an odd number of non-degenerate C-points \cite{Freund2010,Dennis2010,Bauer2015,Galvez2017,Bauer2019,Tekce2019,BAD2019, Wang2021APL}. 

\begin{figure}[t!]
\centering\includegraphics[width = 0.93 \columnwidth]{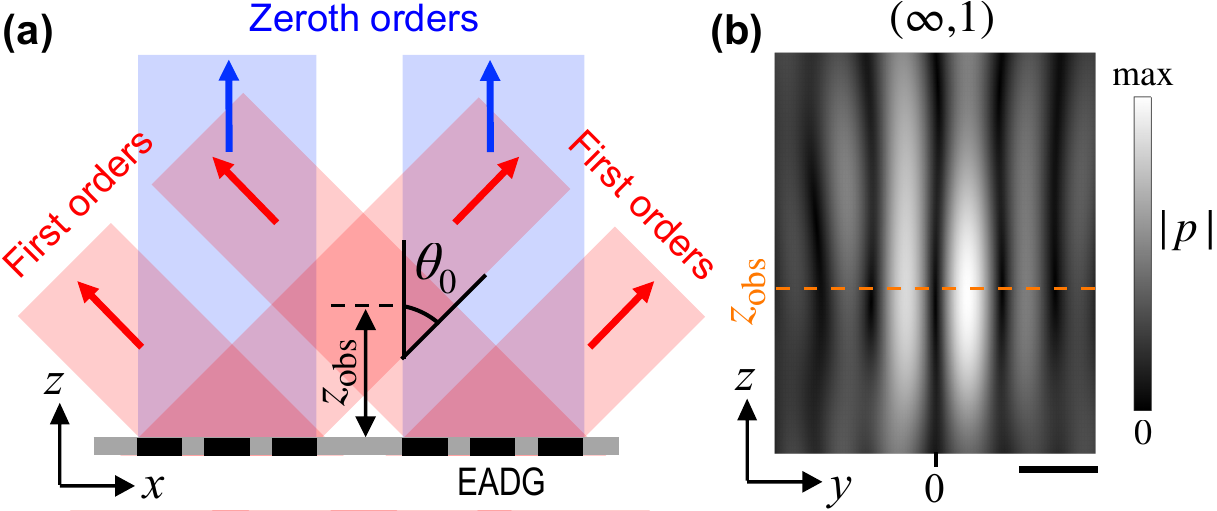}
\caption{(a)  Side view sketch of the diffraction orders emitted from the structured acoustic source. All space-variant transducers are made to operate at $\theta_0=\pi/4$ with the observation plane being located at $\zobs = 22.5$~mm ($ \simeq 6.4\lambda$) for the Bessel beams and $\zobs = 50$~mm for the three-wave interference. (b) Numerically simulated amplitude of the pressure field $p$ in the meridional $(y,z)$ plane for the Bessel beam with $\ell=1$. (For $\ell=-1$ the distribution is mirror symmetric $y \to -y$; the slight asymmetry is due to the broken cylindrical symmetry of the spiral source, Fig.~\ref{Fig1}.) Scale bars: $\lambda$.}
\label{Fig3}
\end{figure}

{\it Experimental setup.---}
Space-variant ferroelectret-based acoustic transducers, which can be regarded as examples of active acoustic metasurfaces \cite{Assouar2018}, are at the heart of the setup shown in Fig.~2. Their principle of operation is detailed in \cite{Ealo2009}. This technology allows operating over a broad ultrasonic frequency range 60--300\,kHz and offers creating space-variant sources on either flat or developable surfaces. Good acoustic impedance matching with air and user-friendly operation make it a valuable experimental option. 

To generate the idealized plane-wave superpositions depicted in Fig.~\ref{Fig1} and given by Eqs.~(\ref{eq1}) and (\ref{eq2}), we fabricate three types of flat monolithic electro-active diffracting gratings (EADGs): counterclockwise and clockwise circular spirals for the Bessel beams with $(N,\ell)=(\infty, \pm1)$ and a clockwise triangular spiral for the three-wave interference $(N,\ell)=(3, -1)$, see Fig.~\ref{Fig2}. Details on the design, fabrication and characterization of  EADGs  for the generation of ultrasonic Bessel beams can be found in \cite{Muelas2018_APL, Muelas2018_JASA}. Here, the space-variant acoustic transducers are excited with a computer-controlled 100~V peak-to-peak chirped signal driving a signal generator (Tektronix, Model AFG3022C), connected to a high-speed high-voltage amplifier (Falco System, Model WMA300) for power supply of the EADGs.

\begin{figure}[t!]
\centering\includegraphics[width = 0.8\columnwidth]{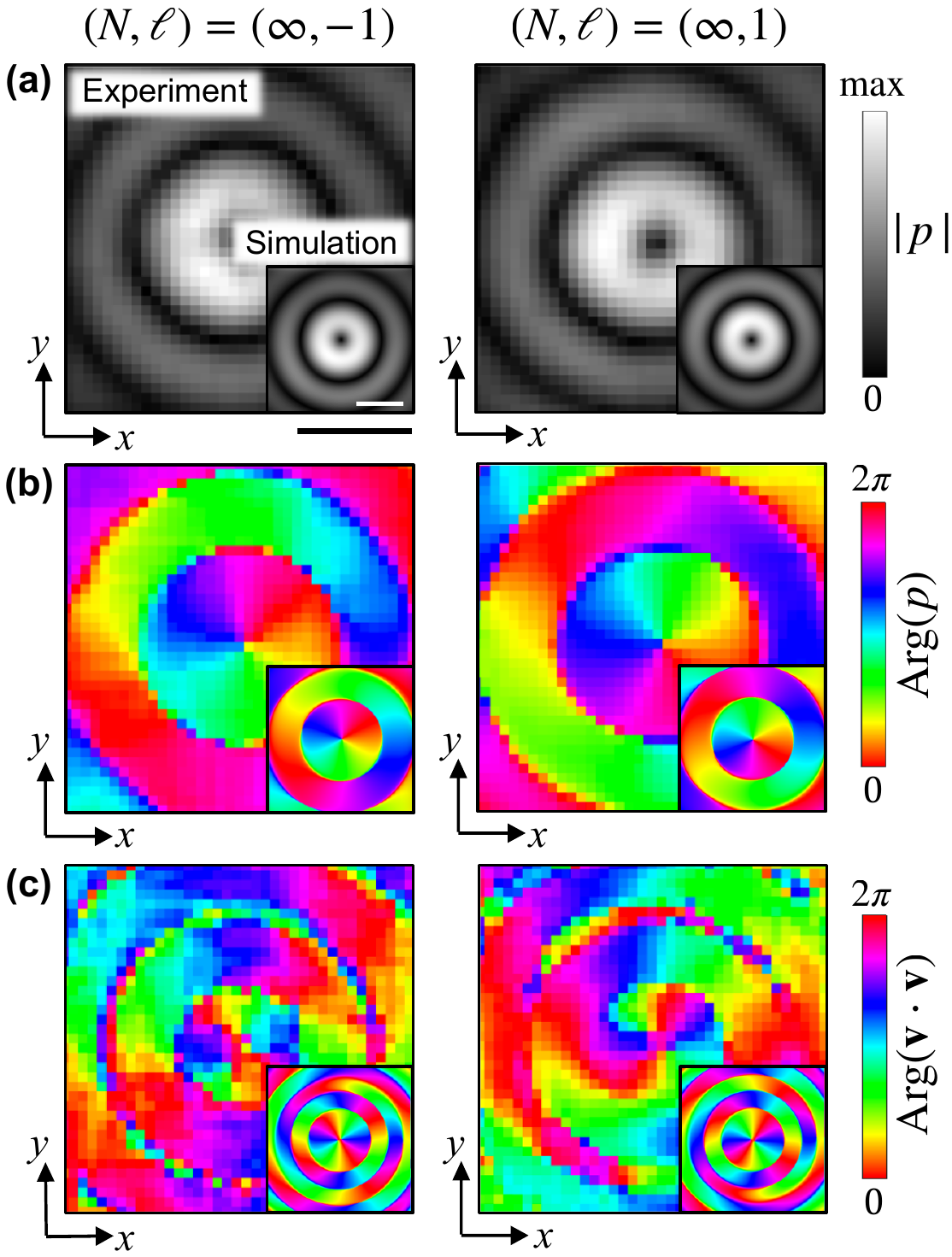}
\caption{Experimental (main panels) and numerically simulated (inset panels) transverse $(x,y)$ distributions (at $z=\zobs$) of the amplitude (a) and phase (b) of the pressure field $p$, and of the phase (c) of the quadratic field $\bfv \cdot \bfv$ for the Bessel beams with $(N,\ell)=(\infty,\pm 1)$. The charge-$\pm 2$ vortices in ${\rm Arg}(\bfv \cdot \bfv)$ indicate double-degenerate C-points of circular polarizations in the beam centers \cite{Bliokh2019_II,Bliokh2021}. Scale bars: $\lambda$.}
\label{Fig4}
\end{figure}

The geometrical grating parameters are introduced in Fig.~\ref{Fig2} (in red), where $dP=0.45 P$. The desired interference field is generated owing to the first-order diffraction from the grating characterized by the angle $\theta_0=\pi/4$ [see Fig.~\ref{Fig3}(a)], which is equivalent to $P=\sqrt{2}\,\lambda$ where $\lambda = 2\pi c/\omega$ is the wavelength. The azimuth-dependent phase difference between the diffracted waves, $\Phi_j = \ell \phi_j$, is ensured by the geometry of the spiral \cite{Muelas2018_APL,Muelas2018_JASA}. 
The circular Bessel-beam spirals have $M=5$ turns with a minimum radius $r_0=10$~mm, and they operate at the frequency $\omega/2\pi = 97$~kHz ($\lambda = 3.5$~mm). The triangular spiral has $M=8$ turns with $r_0=22$~mm (the minimum radial distance to the spiral), and it operates at $\omega/2\pi=70$~kHz ($\lambda = 4.9$~mm). 

In the limit of geometric acoustics the first diffraction orders overlap at axial distances between  
\begin{equation}
\hspace{-2mm}z_{\rm min}=r_0\frac{P}{\lambda} \sqrt{1-\frac{\lambda^2}{P^2}} 
~~{\rm and}~~
z_{\rm max}=z_{\rm min}\!\left(1+\frac{MP}{r_0}\right)
\end{equation}
from the acoustic source. Therefore, we obtain the desired fields at a distance $z_{\rm obs} = (z_{\rm max}+z_{\rm min})/2$ from the source plane, as shown in Fig.~\ref{Fig3}(a). Note that $r_0 \neq 0$ prevents the zeroth-order diffracted field to interfere with the desired first-order one in the $(x,y)$ area $r < r_0$ (there are no propagating higher diffraction orders at the frequencies of interest).

\begin{figure}[t!]
\centering\includegraphics[width = 0.95 \columnwidth]{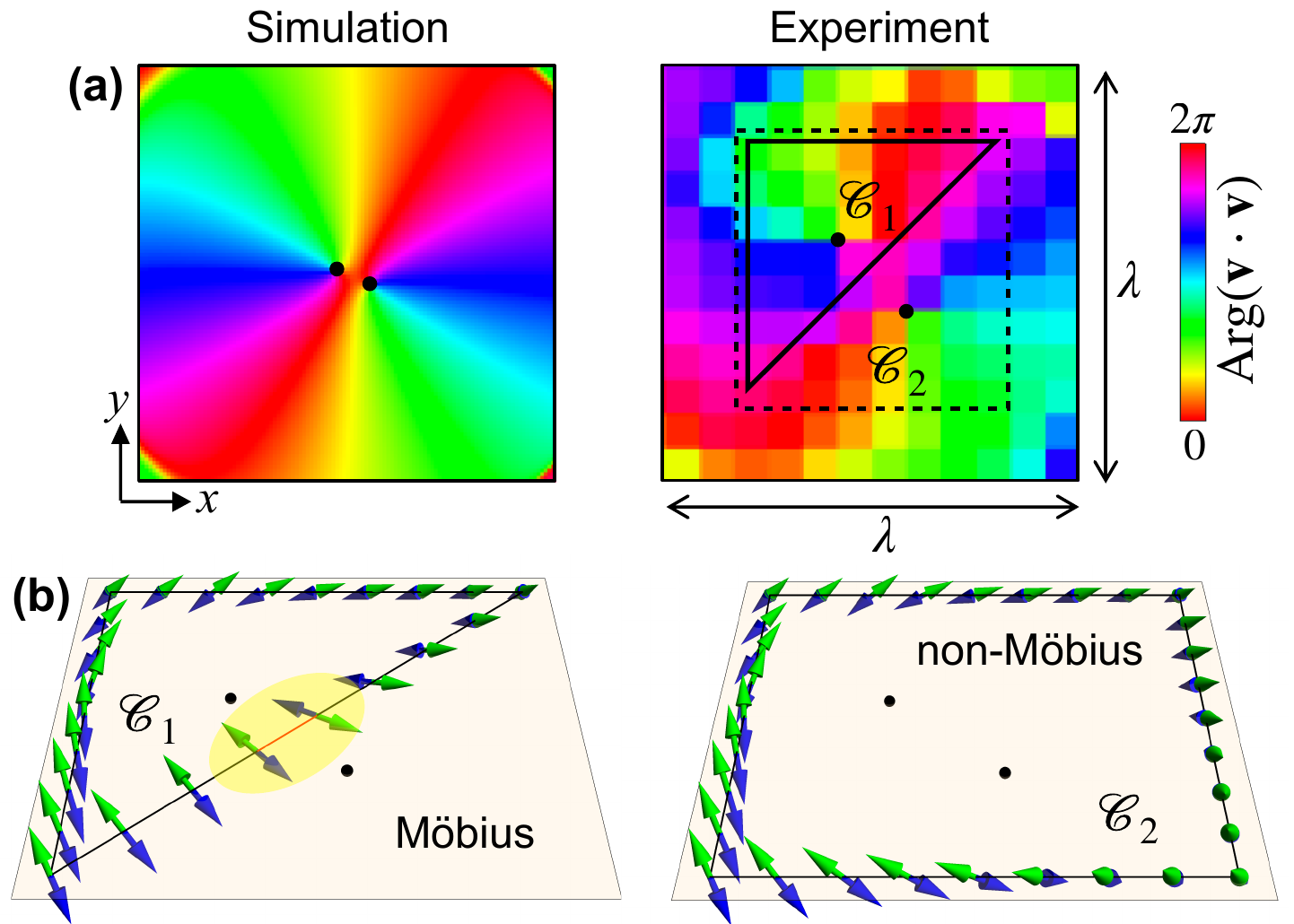}
\caption{(a) Fine subwavelength splitting of the double-degenerate C-point in the center of the Bessel beam with $\ell =1$, Fig.~\ref{Fig4}(c), into a pair of nondegenerate C-points due to the slight asymmetry of the spiral source. (b) Experimentally retrieved distributions of the major semi-axes (represented by colored bi-vectors) of the polarization ellipses of the 3D velocity field ${\bf v}$ along the contours ${\cal C}_1$ and ${\cal C}_2$ embracing one and two C-points, respectively. These two configurations correspond to the {\mobius}-strip (notice the bi-vector discontinuity highlighted in yellow) and non-{\mobius} topologies \cite{BAD2019}. }
\label{Fig5}
\end{figure}

\begin{figure*}[t!]
\centering\includegraphics[width = 2.05\columnwidth]{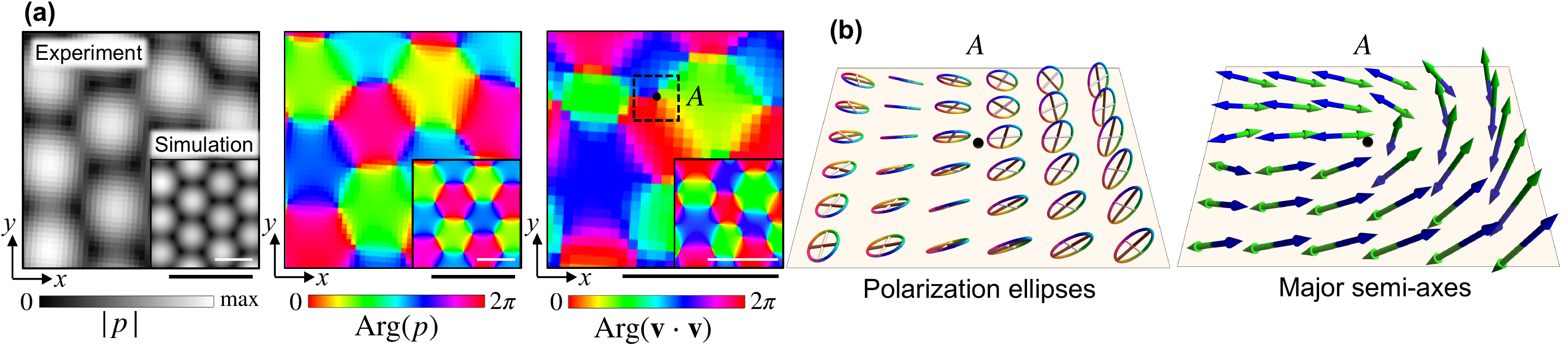}
\caption{(a) Same as Fig.~\ref{Fig4} but for the three-wave interference field $(N,\ell)=(3,-1)$. 
The lattice of charge-$\pm 1$ vortices in ${\rm Arg}(\bfv \cdot \bfv)$ indicate non-degenerate C-points \cite{Bliokh2021}. Scale bars: $\lambda$. (b) Distributions of the experimentally retrieved 3D polarization ellipses and their major semi-axes for the velocity field ${\bf v}$ in the subwavelength area $A$ around one of the C-points. One can see the appearance of the {\mobius}-strip topology around the polarization singularity \cite{Freund2010}.}
\label{Fig6}
\end{figure*}

{\it Measurements and simulations.---}
The \red{polychromatic} radiated acoustic pressure field \red{(50--200~kHz chirped pulse  of  830~$\mu s$ of duration at a repetition rate of 10~Hz)} is measured using a calibrated microphone (1/8 inch, Bruel and Kjaer) whose position is controlled in 3D with an XYZ stage unit. A \red{$0.5~$mm diameter} pinhole is fitted to the tip of the microphone to increase the spatial resolution of the measurements \red{while preventing diffraction drawbacks \cite{Sasaki2006}. The signal is recorded after being amplified while a real-time monitoring is made with an oscilloscope. Then, we perform a Fourier analysis of the time-domain data, which gives access to both the magnitude and phase of each spectral component of the pressure field with practically negligible noise. The frequency of interest (70 kHz or 97 kHz) is selected according to our design for further data processing}.

To reconstruct the 3D inhomogeneous acoustic field, we gather data both in the $(x,y)$ transverse plane, at $z=z_{\rm obs}$, and in the $(x,z)$ meridional plane. Measurements in the transverse plane are made over a squared grid of $12 \times 12$\,mm$^2$ [$15 \times 15$\,mm$^2$] in steps of 0.3 mm corresponding to the $(x,y)$ planes at the distances of $z=(22.8, 23.0, 23.2)$\,mm [$z=(50.3, 50.6, 50.9)$\,mm] for the Bessel beams $(N,\ell)=(\infty, \pm1)$ [three-wave interference $(N,\ell)=(3, -1)$]. These values of $z$ are experimentally chosen as optimal and are close to the expected values $z_{\rm obs}$. The distance between the three consecutive $z$-planes and points of the grid in the $(x,y)$-plane allow one to reconstruct the distribution of the vector velocity field ${\bf v}({\bf r}) \propto {\bm \nabla} p({\bf r})$ in the observation $(x,y)$-plane from the calculated gradient of the scalar pressure field. Similar square grids in the $(x,z)$-planes with $y=(-0.2,0,0.2)$\,mm are used for measurements and reconstruction of the velocity field in the meridional $(x,z)$-plane. For all measurements, the data being shown are evaluated after convolution smoothing of the raw complex pressure field over $5 \times 5$ data points. 

Numerical validation is performed by simulating the generated acoustic fields based on the Rayleigh diffraction integral for the total radiated pressure field, namely,
\be
\label{eq3}
p(\bfr) = -i\frac{\rho \omega}{2\pi} \iint_{S} {v_z}(\textbf{r}_s)\frac{\exp(ik| \bfr - \bfr_s| )}{| \bfr - \bfr_s |}dS\,,
\ee
where ${v_z}(\textbf{r}_s)$ is the vertical velocity of the structured transducer at a point $\textbf{r}_s$ in the $z=0$ plane. As an example, Fig.~\ref{Fig3}(b) shows the pressure field magnitude of the Bessel beam with $\ell=1$ in the meridional plane $(y,z)$.

{\it Results.---}
Figure~\ref{Fig4} shows the results of numerical simulations and experimental measurements of the transverse distributions of the amplitude and phase of the pressure field $p({\bf r})$, as well as of the phase of the quadratic field ${\bf v}({\bf r}) \cdot {\bf v}({\bf r})$ for the Bessel beams with $\ell = \pm 1$ generated by the circular spirals. One can see the charge-$\ell$ vortex at the center of the pressure field $p$ and the charge-$2\ell$ vortex at the center of the ${\bf v} \cdot {\bf v}$ field. The latter shows the double-degenerate C-point at the Bessel-beam center \cite{Bliokh2019_II,Bliokh2021}. A zoomed-in view on the central area, shown in Fig.~\ref{Fig5}(a), reveals the fine splitting of the double-degenerate C-point into a pair of non-degenerate C-points. This splitting occurs because the spiral source is not perfectly cylindrically-symmetric, which lifts the central degeneracy in the sound Bessel beam \cite{Bliokh2021,Bliokh2008,Vyas2013,Otte2018}. 

Having two non-degenerate C-points, we can resolve the polarization M\"{o}bius-strip structure around each of these. Figure~\ref{Fig5}(b) shows the bi-vectors of the major semi-axes of the polarization ellipses for the 3D velocity field ${\bf v}({\bf r})$, retrieved from the experimental data, along the two contours ${\cal C}_1$ and ${\cal C}_2$ embracing one and two C-points. One can see the M\"{o}bius and non-M\"{o}bius evolutions of the polarization orientations along these contours, in agreement with the theory of polarization singularities \cite{Freund2010,BAD2019}. 

Figure~\ref{Fig6} shows analogous results of numerical  simulations and experimental measurements for the three-wave interference with $\ell = - 1$ generated by the triangular spiral. One can see a periodic lattice of charge-$\pm 1$ vortices in the pressure field $p$ and in the ${\bf v} \cdot {\bf v}$ field. Thus, the field exhibits a lattice of non-degenerate C-points \cite{Bliokh2021}. Figure~\ref{Fig6}(b) shows the distribution of the polarization ellipses and their major semi-axis bi-vectors for the 3D field ${\bf v}({\bf r})$ in the vicinity of one of the C-points. One can clearly trace the appearance of the M\"{o}bius-strip topology around the polarization singularity.

\begin{figure}[t!]
\centering\includegraphics[width = 1\columnwidth]{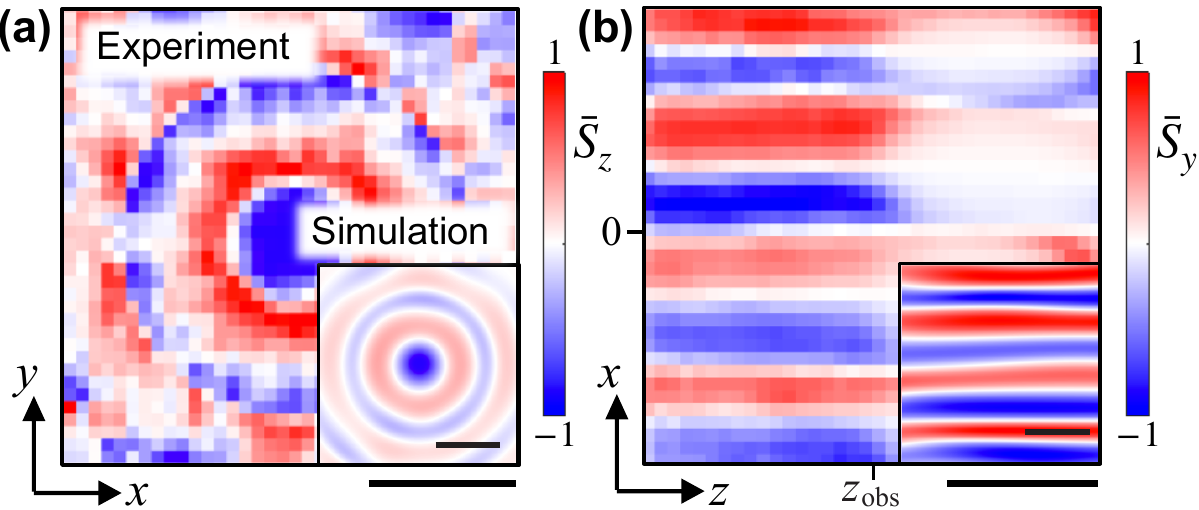}
\caption{Experimentally retrieved and numerically calculated (insets) distributions of the longitudinal $z$ and transverse $y$ components of the normalized spin density $\bar{\bf S}$ in the transverse $(x,y)$ (at $z=\zobs$) and meridional $(z,x)$ (at $y=0$) cross-sections of the Bessel beam with $\ell=-1$. Flipping the sign of $\ell$ flips the sign of $\bar{S}_z$ (see \blue{Supplemental Information, Section 1}) but not $\bar{S}_y$ \cite{Bliokh2019_II}. Scale bars: $\lambda$. }
\label{Fig7}
\end{figure}

It is worth noting that, since we obtain complete information about the vector velocity field ${\bf v}({\bf r})$, we can assess experimentally any polarization-related properties of the inhomogeneous sound wave field. This is illustrated in Fig.~\ref{Fig7} which displays the longitudinal and transverse components of the normalized 3D spin density $\bar{\bf S} = {\rm Im}\left( {\bf v}^* \times {\bf v} \right)/ |{\bf v}|^2$ in the transverse and meridional cross-sections of the Bessel beam with $\ell=-1$, constructed from the measurements. These experimental results, which are supported by numerical simulations, validate recent analytical calculations \cite{Bliokh2019_II}.

Also, we find that the distribution of the direction of the instantaneous velocity field ${\rm Re}\!\left[{\bf v}({\bf r})e^{-i\omega t}\right]$ in the three-wave interference experiment exhibits a lattice of {\it skyrmions} \cite{Beckley2010,Donati2016,Tsesses2018,Du2019,Gao2020,Sugic2021}, \red{see Fig.~8 and} \blue{Supplemental Information, Section 2}). This echoes the recent results of Ref.~\citenum{Ge2021} but for propagating acoustic waves rather than for surface standing waves. Moreover, the temporal behavior of the skyrmionic pattern and the corresponding Skyrme number are discussed in the Supplementary Information, both for the theoretical and the experimental cases.

{\it Conclusions.---}
We presented experimental measurements of polarization singularities (C-points) and 3D topological polarization M\"{o}bius-strip structures in inhomogeneous sound-wave fields. In doing so, we employed 2D spiral gratings generating nonparaxial interference fields with imprinted vortex properties. By changing the approximate rotational symmetry of the spirals (triangular and circular in our experiments) we controlled the symmetry properties of the generated propagating fields, demonstrating near-degenerate and non-degenerate polarization singularities. In all cases, evolution of the 3D polarization ellipse of the vector velocity field along a contour embracing an odd number of C-points exhibits the M\"{o}bius-strip topology, in agreement with the general theory of polarization singularities. We also retrieved nontrivial distributions of the 3D spin density in nonparaxial sound Bessel beams, as well as skyrmionic features in the instantaneous velocity field for a superposition of three plane waves. \red{All generic topological structures we observed are robust against small perturbations, which can be seen from our experimental data exhibiting inevitable distortions.} Our results pave the way to further investigations and applications of topological polarization features of \red{3D vector} waves. \red{In addition to the sound waves considered here, one could study other kinds of material waves, such as water waves, elastic waves, etc.}
 
\vspace{2mm}
This work was partially funded by project PAPIIT-IN113422 by DGAPA-UNAM, M\'exico. R.D.M-H acknowledges support from Colciencias Scholarship program No. 727.  F.N. acknowledges partial support from NTT Research. M.A.A. and E.B acknowledge support from the French National Research Agency (ANR) through award ANR-21-CE24-0014-01.
\vspace{-4mm}
\begin{figure}[t!]
\centering\includegraphics[width=0.95\columnwidth]{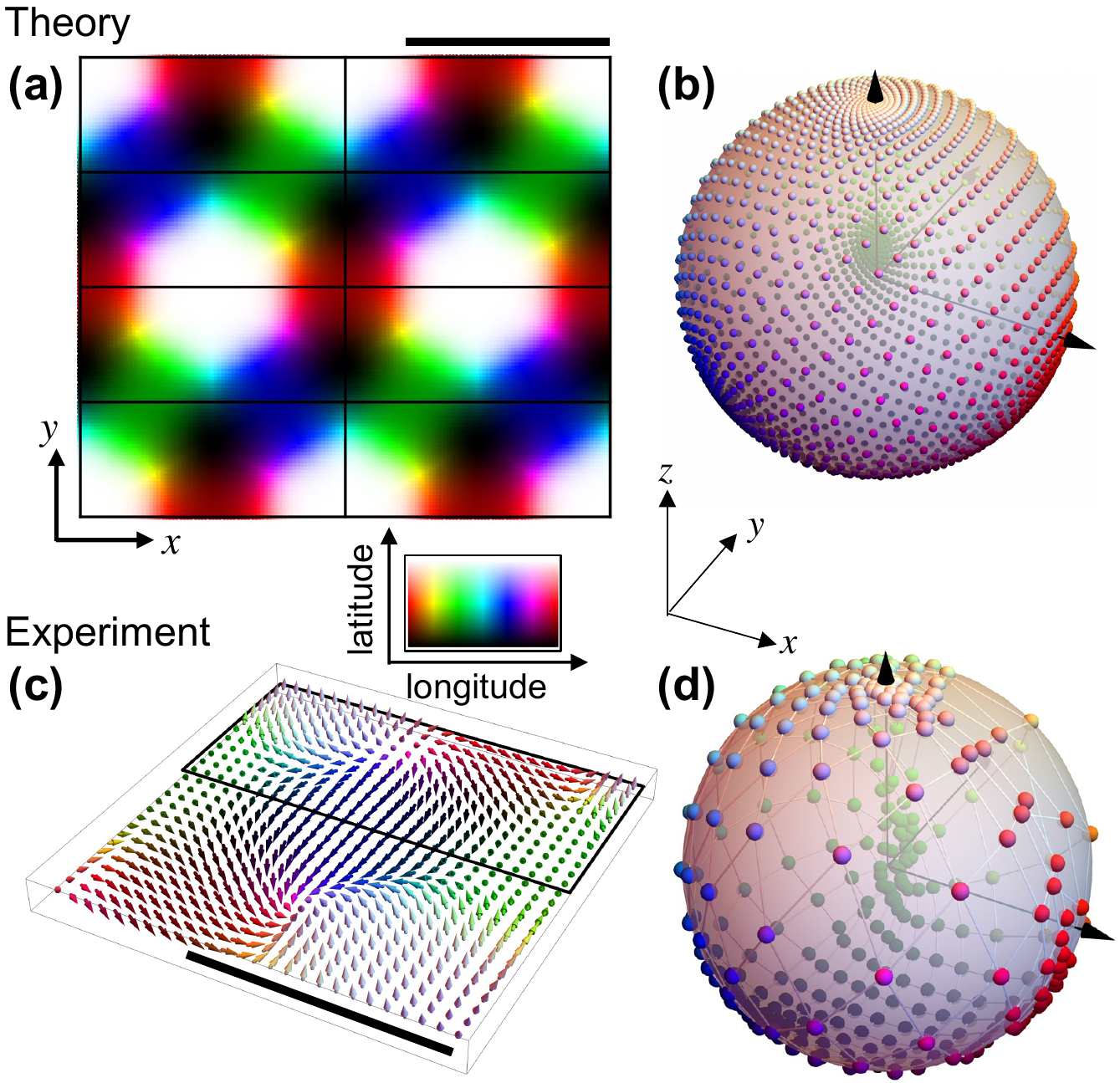}
\caption{
\red{Theoretical (top row) and experimental (bottom row) coverage of the sphere of velocity field directions for a superposition of three plane waves. In all parts, the direction of the velocity field is encoded as color following the palette at the center. (a) Theoretical distribution of directions for the velocity field over the transverse $(x,y)$ plane, which can be subdivided into cells in which each direction is covered once. (b) Coverage of the sphere of directions for a Cartesian sampling of the $(x,y)$ plane. (c) Velocity field direction distribution for the measured field. (d) Distribution over the sphere of directions for the cell indicated by a rectangle in (c). Scale bars: $\lambda$. See also} \blue{Supplementary Videos 1 and 2}.}
\end{figure}

\newpage
\begin{center}
{\bf SUPPLEMENTAL INFORMATION}\\
\end{center}

\textbf{1. $\ell$-dependent ${\bar S_z}$ for Bessel vortex fields}
\vspace{2mm}

The figure S1 shows experimental and simulated dependence of the longitudinal component of the normalized spin density, ${\bar S_z}$, on the topological charge $\ell$ of Bessel vortex beams, which has been previously predicted analytically in Ref.~\citenum{Bliokh2019_II}. According to the notation introduced in the main text, the results refer to $(N,\ell) = (\infty,\pm1)$

\begin{figure}[h]
\centering\includegraphics[width=1\columnwidth]{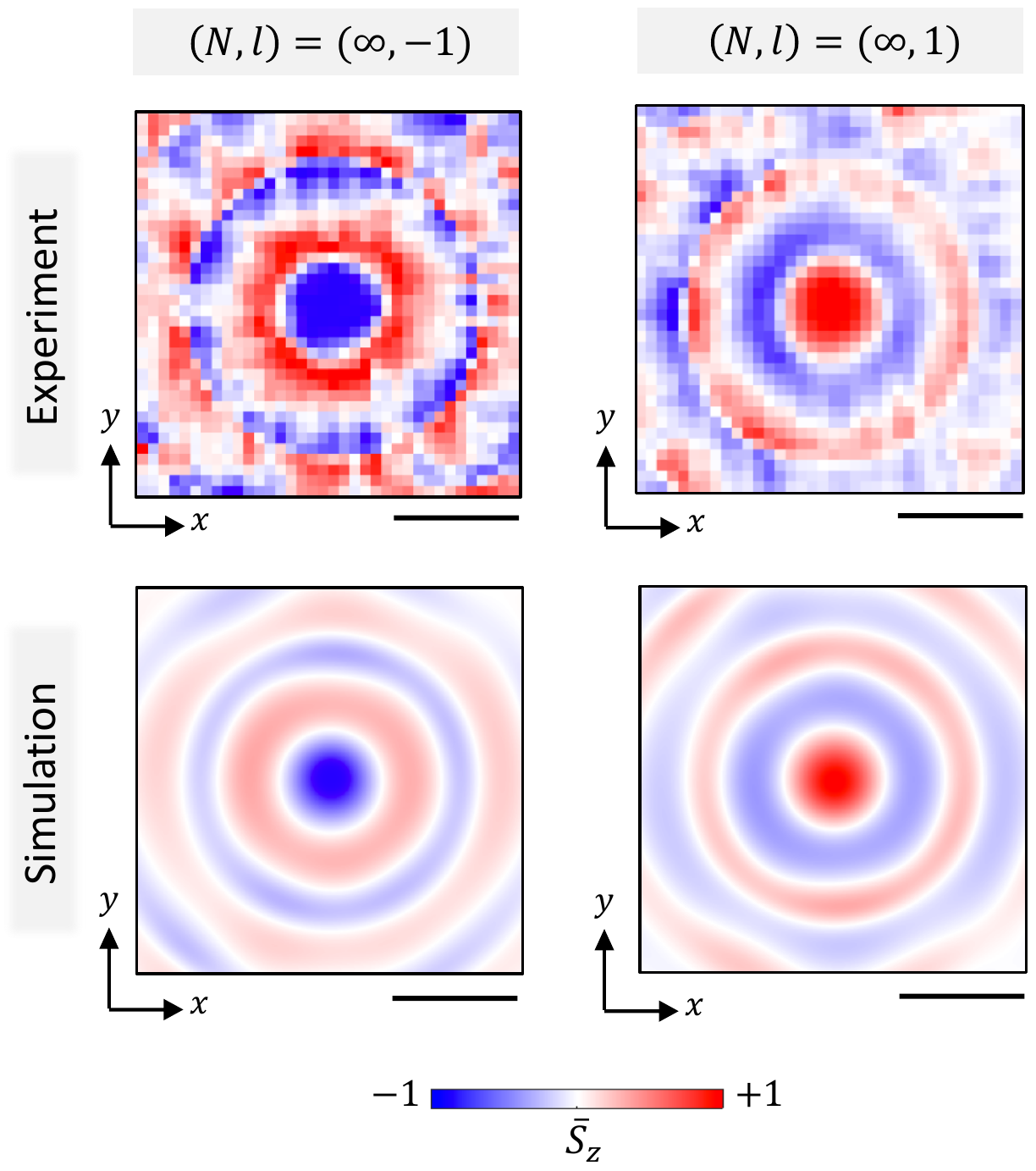}
\caption{
Experimentally retrieved (top row) and numerically calculated (bottom row) distributions of the longitudinal $z$ component of the normalized spin density $\bar{\bf S}$ in the transverse $(x,y)$ (at $z=\zobs$) for Bessel beams with $\ell=\pm1$. Scale bars: $\lambda$. 
}
\end{figure}

\noindent
\textbf{2. Skyrmion texture in three-wave interference} 
\vspace{2mm}

Here we show that, for fixed $t$ and $z$, the direction of the velocity field for a superposition of three plane waves covers the complete sphere when the coordinates ($x,y$) vary within a given area. The velocity field at this time can then be regarded as a skyrmionic lattice. 

The velocity field distribution at, say, $t=0$ and $z=0$ (where the field at the origin is maximal) corresponds to the real part of Eq.~(2), in this case with $N=3$ and $\Phi_j=0$, that is, $\ell=0$. Recall that, while the experimental implementation used $\ell=-1$, for $N=3$ a change in $\Phi_j$ corresponds to a spatial shift of the pattern and a global phase.  Figure~8(a) of the main text shows a simulation based on Eq.~(2) of the direction of the velocity field at each point, with color-encoded direction: hue encodes the azimuthal coordinate (longitude) and light-dark encodes the polar coordinate (latitude). The pattern in Fig.~8(a) can be subdivided into cells over which all directions appear only once. This subdivision is not unique, but for convenience we choose aligned horizontal rectangular sections. Note that the vertical limiting lines can be shifted arbitrarily without changing the fact that each direction is covered only once, while the horizontal lines are fixed. (In fact, the vertical boundaries of the third and fourth rows could be shifted by half a period to better match the patterns of the first and second rows, but such change is irrelevant to the covered velocity field directions.) The coverage of the sphere of directions for a Cartesian sampling of one of these cells is shown in Fig.~8(b) and \blue{Supplementary Video 1}.

The measurements follow the theoretical predictions. Figure~8(c) shows the skyrmionic texture for the measured velocity field at $t=0$, using the same data over the same area as that in the third panel in Fig.~6(a). The top section (enclosed in a rectangle) approximately corresponds to a cell like those in Fig.~8(a) (with the vertical boundaries shifted). The coverage of directions over the sphere for the measured points is shown in Fig.~8(d) and \blue{Supplementary Video 2}.

\red{We now discuss how the coverage of the sphere of directions changes with time. At any given time $t$ and distance $z$, the density of the mapping from the plane $(x,y)$ to the surface of the sphere of directions is characterized by the Skyrme density, defined as
\be
    \sigma(x,y;z,t)=\frac1{4\pi}\hat{\bf u}\cdot\frac{\partial\hat{\bf u}}{\partial x}\times\frac{\partial\hat{\bf u}}{\partial y}, 
\ee
where
\be
   \hat{\bf u}=\frac{{\rm Re}\!\left[{\bf v}({\bf r})e^{-i\omega t}\right]}{|{\rm Re}\!\left[{\bf v}({\bf r})e^{-i\omega t}\right]|}. 
\ee
The integral of the Skyrme density over a unit cell gives the Skyrme number, 
\be
    \Sigma(z,t)=\iint_{\rm unit\,cell} \!\!\!\sigma(x,y;z,t)\,dx\,dy,
\ee
which equals $\pm1$ if the sphere is fully covered over the corresponding area, the sign determining the sense in which the coverage takes place.}

\red{As Fig.~8 shows, for $t=0$ and $z=0$ the coverage is fairly uniform. However, for other times (or equivalently, for other propagation distances $z$) it can become more irregular, and in fact the sign of the Skyrme number presents abrupt changes. 
The reason for this can be appreciated from \blue{Supplementary Video 3}, which shows the evolution of the coverage over a temporal period, as well as the skyrmionic texture. One can see that there are topological transitions at certain times and locations (associated with zeros of the field). The video also shows that the even and odd rows of cells shown in Fig.~8(a) no longer cover each the complete sphere of directions separately; for each some sections are not covered and others are covered twice. However, together such two contiguous cells do cover the complete sphere twice.   
Note that for $t=mT/6$, where $m$ is an integer and $T=2\pi/\omega$ is the temporal period of the acoustic wave, the coverage of the sphere is similar to that of $t=0$, where each cell covers fairly uniformly the whole sphere.}

\red{
Figure S2 shows the temporal evolution of the Skyrme number averaged over two vertically contiguous cells (see Fig.~8), $\overline{\Sigma}$, both for the theoretical case (black curve) and for the experimental data (blue curve). For the theoretical case the coverage is complete almost at any time, but it switches sign six times per cycle. The experimental counterpart presents similar oscillations, however without exhibiting a perfect square waveform due not only to experimental limitations but also  to the coarse sampling that results in a rough estimations of the spatial derivatives and of the integral.}\\

\begin{figure}[t!]
\centering\includegraphics[width=1\columnwidth]{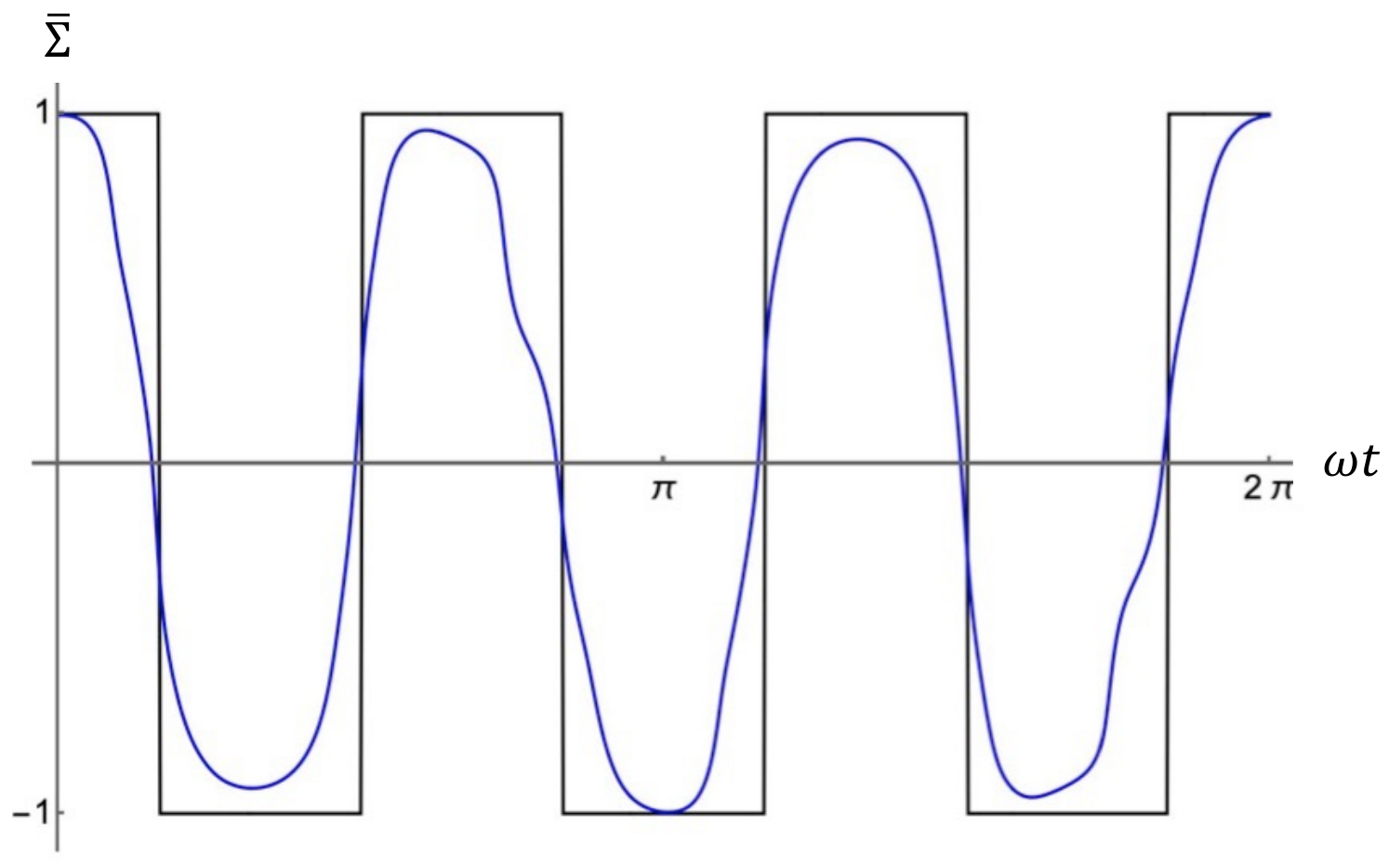}
\caption{
\red{Skyrme number $\overline{\Sigma}$  averaged over two vertically contiguous cells (see Fig.~8), as a function of time over a wave cycle at fixed $z$, for the theoretical superposition of three plane waves (black curve) and for the experimental measurements (blue curve).}
}
\end{figure}

\noindent
\textbf{3. Video captions} 
\vspace{3mm}

{\bf Video 1.} For a theoretical model of three plane waves: (left) coverage of the sphere of directions; (middle) color representation of distribution of directions over the $(x,y)$ plane; (right) skyrmionic texture (at one third of the sampling of the other two parts). In all parts, the color scheme for representing velocity field directions follows the palette at the center of \red{Fig.~8.}
\vspace{2mm}

{\bf Video 2.} For the experiment using a triangular grating: (left) coverage of the sphere of directions, where due to the lower sampling a grid of white lines is used to aid visualisation; (middle) color representation of distribution of directions over the $(x,y)$ plane; (right) skyrmionic texture. In all parts, the color scheme for representing velocity field directions follows the palette at the center of \red{Fig.~8.}, and the sampling corresponds to that of the experimental data.
\vspace{2mm}

{\bf Video 3.} \red{Time evolution of the velocity field directions for a theoretical model of three plane waves. The distributions of points over the spheres at the top and bottom panels on the left column correspond, respectively, to the coverage of velocity field directions of the cells over the first and second (or equivalently third and fourth) rows of any of the two columns of the plane in the top-right panel.} The bottom-right panel shows the skyrmionic texture corresponding to the two cells at the top-left of the panel above.


\end{document}